\documentclass[10pt]{bmc_article}

\usepackage{graphicx,color}
\usepackage{amsmath,amssymb}
\usepackage{cite} 
\usepackage{url}  
\usepackage{ifthen}  
\usepackage{multicol}   
\usepackage[utf8]{inputenc} 
\newboolean{publ}

\newenvironment{bmcformat}{\begin{raggedright}\baselineskip20pt\sloppy\setboolean{publ}{false}}{\end{raggedright}\baselineskip20pt\sloppy}

\begin{document}
\begin{bmcformat}


\title{Comparative study of nonlinear properties of EEG  signals of normal persons and  epileptic patients}


\author{Md. Nurujjaman\correspondingauthor$^{1}$%
       \email{Md. Nurujjaman\correspondingauthor - jaman\_nonlinear@yahoo.co.in}%
      \and
         Ramesh Narayanan$^2$%
         \email{Ramesh Narayanan- rams@plasma.inpe.br}
       and
         A.N. Sekar Iyengar$^1$%
         \email{A.N. Sekar Iyengar - ansekar.iyengar@saha.ac.in}%
      }


\address{%
    \iid(1)Plasma Physics Division, Saha Institute of Nuclear Physics,
     1/AF, Bidhannagar, Kolkata -700064, India\\
    \iid(2)Laboratorio Associado de Plasma, Instituto Nacional de Pesquisas Espaciais, Av. dos Astronautas, 1758 - Jardim da Granja 12227-010 Sao Jose dos Campos, SP, Brazil
    }%

\maketitle


\begin{abstract}
        \paragraph*{Background:} Investigation of the functioning of the brain in living systems has been a major effort amongst scientists and medical practitioners. Amongst the various disorder of the brain, epilepsy has drawn the most attention because this disorder can affect the quality of life of a person. In this paper we have reinvestigated the EEGs for normal and epileptic patients using surrogate analysis, probability distribution function and Hurst exponent.

        \paragraph*{Results:} Using random shuffled surrogate analysis, we have obtained some of the nonlinear features that was obtained by Andrzejak \textit{et al.} [Phys Rev E 2001, 64:061907], for the epileptic patients during seizure. Probability distribution function shows that the activity of an epileptic brain is nongaussian in nature. Hurst exponent has been shown to be useful to characterize a normal and an epileptic brain and it shows that the epileptic brain is long term anticorrelated whereas, the normal brain is more or less stochastic. Among all the techniques, used here, Hurst exponent is found very useful for characterization different cases.
        \paragraph*{Conclusions:} In this article,  differences in characteristics for normal subjects with eyes open and closed, epileptic subjects during seizure and seizure free intervals have been shown mainly using Hurst exponent. The H shows that the brain activity of a normal man is uncorrelated in nature whereas, epileptic brain activity shows long range anticorrelation.
\end{abstract}

\ifthenelse{\boolean{publ}}{\begin{multicols}{2}}


\section{Background}
The brain is a highly complex and vital organ of a human body whose neurons  interact with the local as well as the remote ones in a very complicated way~\cite{PRE:Andrzejak,PRE:Diambra,PLA:Rombouts,chaos:Mathew}. These interactions evolve as the spatio-temporal electro magnetic field of the brain,  and are recorded as Electroencephalogram (EEG) ~\cite{chaos:Mathew,PRE:Andrzejak,PRE:Robinson,ClinicalNeuro:stam}. Though the detail link between EEGs and the underlying physiology is not well understood, the former is widely used for detection and prediction of epilepsy, localization of epileptic zone and characterization of the pre and post-ictal~\cite{PRE:Andrzejak,ClinicalNeuro:stam,PRE:pradhan} using linear and nonlinear analysis techniques~\cite{PRE:Andrzejak,ClinicalNeuro:stam,PRE:pradhan,Neurocomputing:Anna,NeurosciLett:Pereda,NeurosciLett:Pereda1,EpilepsyBehavior:Hughes}.
Though mainly nonlinear methods have been applied to predict the onset of epileptic seizure and localizing epileptic regions,limited progress has been achieved so far~\cite{EpilepsyBehavior:Hughes}. Even some negative results have also been reported like linear measures are better than nonlinear measures~\cite{Clin Neurophysiol:Mormann,JClin Neurophysiol:Jerger}, seizure is not a low dimensional process~\cite{IEEE Trans Biomed Eng:Lopes}, it lacks determinism~\cite{Ann Biomed Eng:Slutzky,JClinNeurophysiol:Savit,Neurocomputing:Anna}, etc. Hence finding proper analysis techniques is also one of the main issues and experts try out different analysis tools for characterizing the normal and diseased brain states, especially the epileptic brain.

In 2001, Ralph G. Andrzejak, \emph{et al.} and later some other authors~\cite{pre:Temujin,physicaD:Harikrishnan} have analyzed five sets of EEG signals~\cite{web:eeg} each set containing 100 epochs to study the determinism in the brain dynamics for five different physiological and pathological conditions. Sets A  and B are for normal persons with eyes open and closed respectively and recorded extracranially. Sets C and D were recorded intracranially from the hippocampal formation which was nonepiletogenic of the opposite hemisphere of the brain and from within the epileptogenic zone  of an epileptic patient during seizure free intervals respectively. Set E was recorded intracranially from the epileptic zone during seizure. The details of the experiments and the conditions have been described in Ref~\cite{PRE:Andrzejak}. R.G. Andrzejak, \emph{et al.}~\cite{PRE:Andrzejak} had shown that the normal healthy subject with eyes closed and open shows stochastic behavior using amplitude adjusted Fourier transform surrogate analysis where discriminating statistics were the effective correlation dimension and nonlinear prediction error whereas, using delay vector variance discriminating statistics, significant nonlinear determinism was shown in the same subject~\cite{pre:Temujin}. So two conflicting results were obtained for the same subject using nonlinear methods. In the case of epileptic patients during seizure and seizure free intervals, determinism was shown using two different methods~\cite{PRE:Andrzejak,pre:Temujin} though other studies show lack of determinism for different epileptic patients during seizure~\cite{Clin Neurophysiol:Mormann,BrainTopogr:Pijn,AnnBiomedEng:Slutzky, JClinNeurophysiol:Savit}.

On the other hand, characterization of EEGs by scaling properties of the signal is also a major area of research interest~\cite{Neurocomputing:Anna,NeurosciLett:Pereda,NeurosciLett:Pereda1,ClinicalNeurophysiology:shen,ClinicalNeurophysiology:Lee,ComputerMethodsandProgramsinBiomedicine:kannathal,ClinicalNeurophysiology:Freeman,pre:robinson1,JournalNeuroscienceMethods:nikolic,BioMedicalEngineeringOnLine:Kannathal1}.
Power spectral exponent has been used to characterize the different subjects with different physiological conditions~\cite{Neurocomputing:Anna,NeurosciLett:Pereda,ClinicalNeurophysiology:Freeman,pre:robinson1} and the same exponent has also been used to estimate the correlation dimension ($D_{corr}$)~\cite{Neurocomputing:Anna}. Fractal dimension and hurst exponent have also been used to characterize the EEGs~\cite{JournalNeuroscienceMethods:nikolic,BioMedicalEngineeringOnLine:Kannathal1}. Hence a number of experts prefer scaling properties to characterize EEG for different physiological and pathological conditions~~\cite{Neurocomputing:Anna}.

In this paper, we have reinvestigated the EEG data studied in Refs.~\cite{PRE:Andrzejak,pre:Temujin,physicaD:Harikrishnan} by random shuffled surrogate analysis using $D_{corr}$ as discriminating statistics in order to find determinism in the signal~\cite{physicaD:theiler,Bifur:Nakamura,ieee:small} and the results have been compared with earlier analyses~\cite{PRE:Andrzejak,pre:Temujin}.
Probability distribution function shows a difference between normal and epileptic brain states and this has been discussed in latter Section. Finally, we have quantified the five different physiological brain states by Hurst exponent (H) which has been estimated using $R/S$ analysis~\cite{POP:carreras1}.

\section{Results and Discussion}
\label{Sec:results}
\subsection{surrogate analysis}
Surrogate analysis  determines the dynamics in the time series: whether it is governed by stochastic or deterministic process~\cite{physicaD:theiler,Bifur:Nakamura,ieee:small}.

\begin{figure}
\center
\includegraphics[height=6cm]{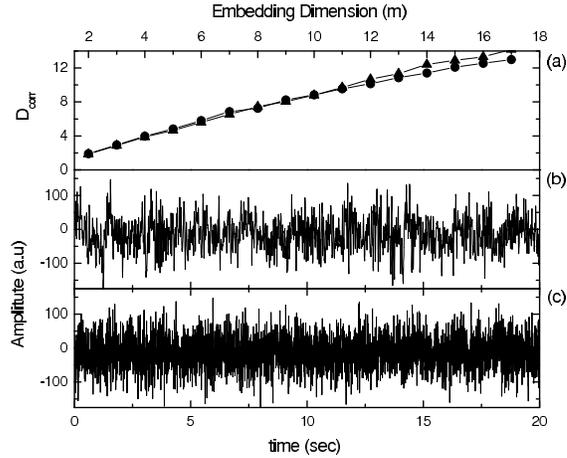}
\caption{surrogate for data Set A: (a) Variation of $D_{corr}$ with m. $\bullet$ for (b) Original data; and $\Delta$ for (c) Surrogate data. }
\label{fig:surgA}
\end{figure}

The surrogate data has been generated by Random Shuffled (RS) surrogate method, in which the signals were shuffled randomly so that the probability distribution is same but the temporal correlations are destroyed~\cite{chaos:dori,physicaD:theiler,Bifur:Nakamura}. $D_{corr}$ which gives us a measure of the complexity has been estimated for both the original and the surrogate data of the data sets A, B, C, D, and E respectively. Fig~\ref{fig:surgA}(a) shows that the $ D_{corr}$ increases with the same trend for both the original [Fig~\ref{fig:surgA}(b)] and the surrogate data [Fig~\ref{fig:surgA}(c)] for the normal persons with eyes open [Set A]. A similar trend is observed in the case of the persons with eyes closed, i.e., for Set B [Fig~\ref{fig:surgBE}(a)]. This shows that the brain activity of a normal person is stochastic in nature agreeing with the analysis by Andrzejak, \emph{et al.}~\cite{PRE:Andrzejak}. For an epileptic patient we found a different behavior during seizure free intervals and during seizure activity. The EEG signals, at seizure free state, recorded both from the hippocampal formation  of the opposite hemisphere of the brain [Fig~\ref{fig:surgBE}(b) Set C] and within the epileptogenic zone [Fig~\ref{fig:surgBE}(c) Set D] show almost same trend in the increase in $D_{corr}$ for the surrogate data as well as for original data, except for a small separation at higher embedding dimensions, which may be due to a high dimensionality of the system.  During seizure activity [Fig~\ref{fig:surgBE}(d) Set E], $D_{corr}$ saturates with embedding dimension indicating low dimensional deterministic dynamics and these results agrees well with previous analyses~\cite{PRE:Andrzejak,pre:Temujin}. For sets A-D, as there is no saturation  in $D_{corr}$ at higher embedding dimension [Fig~\ref{fig:surgA}(a) and Figs~\ref{fig:surgBE}(b)-(d)] and hence it is difficult to estimate actual $D_{corr}$. In Ref~\cite{PRE:Andrzejak} $D_{corr}$ was computed based on quasiscaling regions, but such an estimation is very much dependent on the variations of the time-frequency-energy characteristics rather than any nonlinear dynamics. Hence this may be inadequate to characterize epilepsy or diseased brain states for clinical application~\cite{chaos:Harrison}.

\subsection{probability distribution functions}
\label{subsection:pdf}
As we have observed from the surrogate analysis that nonlinear dynamics is responsible for epileptic patients during seizure, we have compared the probability distribution function (PDF) of a normal case and an epileptic person during seizure. The PDF for sets A and E have been shown in Figs~\ref{fig:pdf}(a) and (b) respectively. Fig~\ref{fig:pdf}(a) shows that for a normal healthy person with eyes open, the PDF is Gaussian in nature, whereas for epileptic patients during seizure, it is nongaussian [Fig~\ref{fig:pdf}(b)] signifying an intermittent nonlinear effect. But for other three cases this feature is not so clear. So we feel that the PDF may also be useful to differentiate a  brain activity of an epileptic patient during seizure from other state.

\begin{figure}
\center
\includegraphics[height=6cm]{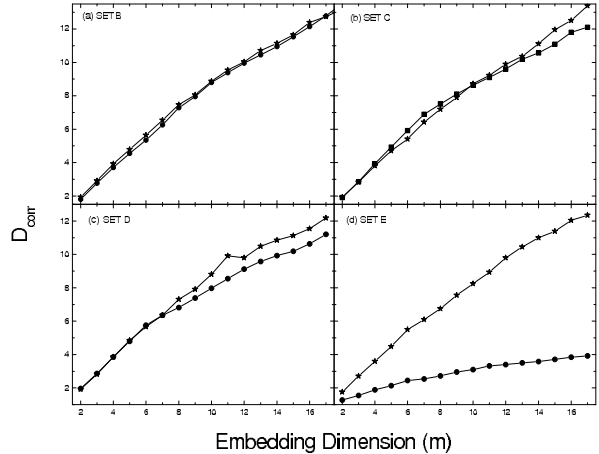}
\caption{$-\bullet-$ for original data and $-*-$ for surrogate data for data Set B, C, D and E}
\label{fig:surgBE}
\end{figure}

\begin{figure}
\center
\includegraphics[height=6cm]{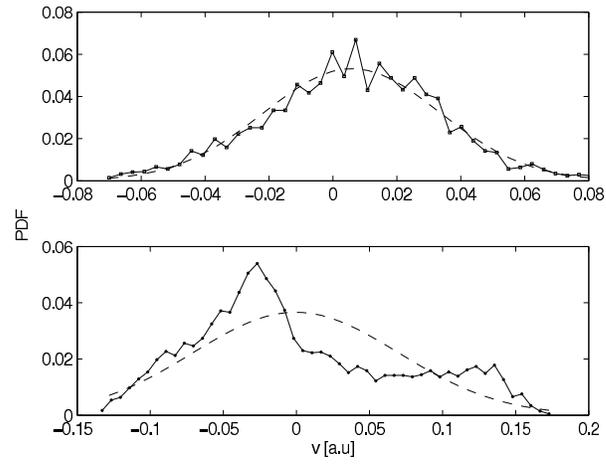}
\caption{Typical PDF for normal person with eyes closed (up) and for epileptic patient during seizure (bottom).}
\label{fig:pdf}
\end{figure}

\begin{figure}
\center
\includegraphics[height=6cm]{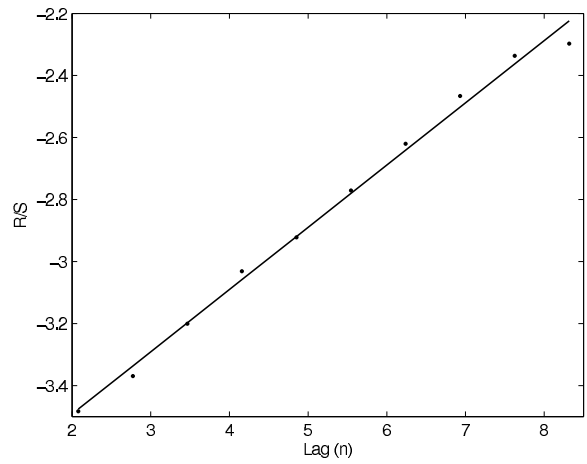}
\caption{Typical R/S vs lag n plot.}
\label{fig:hurst}
\end{figure}

\subsection{Hurst exponent}
\label{subsection:H}
Since one of the major emphasis of epilepsy investigation is to predict their occurrence, it is necessary to know how the data is correlated. We have carried out a study of the Hurst exponent (H) which has been estimated using Rescaled range analysis ($R/S$). This method was proposed by Hurst and well established by Mandelbrot, and Wallis~\cite{POP:carreras1}. For a given set of data series, $R/S$ is defined as~\cite{POP:carreras1,PhyslettA:jaman}:
\begin{eqnarray}
\frac{R(n)}{S(n)}=\frac{max(0,W_1,...,W_n)-min(0,W_1,...,W_n)}{\sqrt{S^2(n)}}
\label{eqn:RS}
\end{eqnarray}
Here $W_k= x_1+x_2+x_3+...+x_k-k\overline{X(n)}$, where $\overline{X}$, $S^{2}(n)$, and n are respectively the mean, variance, and time lag of the signal. The expected value of the $R/S$ scales like $cn^H$ as $n\rightarrow \infty$, where H is called the Hurst exponent, and can be estimated from the slope of typical plot $\frac{R(n)}{S(n)}$ vs lag (n). For a given signal, we divided the data into nonoverlapping blocks of equal length and $R/S$ has been calculated using the Equation~\ref{eqn:RS} and the average value of $R/S$ has been plotted as a function of lag in a $\log-\log$ plot as shown in Fig~\ref{fig:hurst} and estimated H from the slope of the curve.  For random data $H=0.5$, while $H>0.5$  for the data with long range correlations, and $H<0.5$ indicates the presence of long-range anticorrelation or antipersistency in the data.

The estimated  average Hurst exponent ($<H>$) with an error bar of 100 epochs for all the five EEG data sets (viz. A$-$E) have been shown in Fig~\ref{fig:AvgH}. The solid box ($\Box$) and bullet ($\bullet$) show the $<H>$ for sets A and B respectively. For a normal  person with eyes open (set A), the average H ($<H>$) $\approx0.47$ whereas, for the data set B, i.e., for a normal person with his eyes closed, $<H>\approx 0.41$.

\begin{figure}
\center
\includegraphics[height=6cm]{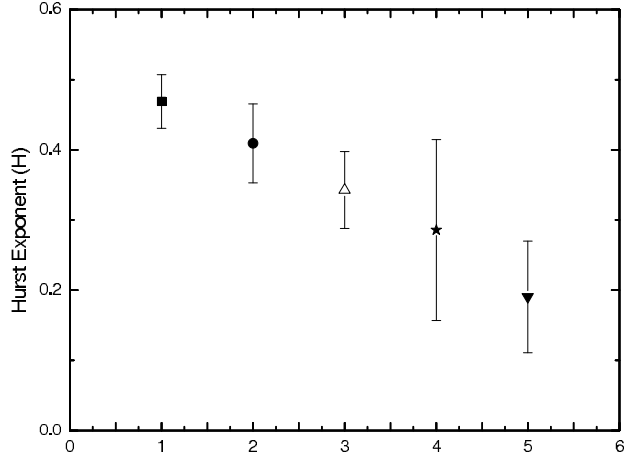}
\caption{Average H with standard deviation error bar, Hurst exponent for hundred time series and average H are represented by: $\Box$  for normal men with eyes open [set A]; $\bullet$ for normal men with eyes closed [set B]; $\triangle$ and $\star$  are for the epileptic patients during seizure free interval from two different locations [ set C and D];$\nabla$  for the epileptic patients during seizure [set E]. }
\label{fig:AvgH} 
\end{figure}

The $<H>\approx 0.5$ for a normal man with eyes open indicates that the signals are uncorrelated over long time scales signifying stochasticity of the normal brain. But with eyes closed state, decreases in H ($<H>\approx0.41$) may be due to the imposition of some extra constraint, which may influence the system towards an antipersistent state. The $<H>$ for epileptic patients  are shown by up triangle ($\triangle$) for set C; star ($\star$) for set D; and down triangle ($\nabla$) for set E respectively. $<H>\approx0.34$ and 0.29 for the EEGs recorded at the hippocampal formation and epileptic zone for the seizure free intervals of epileptic patients respectively and during seizure (data set E), we get the lowest H ($<H>\approx 0.19$). The H for epileptic patient during seizure and seizure free intervals show anticorrelation which may be due to epileptiform discharges during seizure free intervals indicating that a large discharge is always followed by a small one. The physiology behind the epileptiform discharge is due to the chronic dysfunction or ``defect" in the epileptic brain, i.e., the epileptic brain is not normal even during seizure free time~\cite{physicaA:Cajueiro}. Though the hippocampus was nonepileptogenic for these subjects, its H is still less than a normal person which may be due to its participation in secondary, nonautonomous epileptic processes initiated by the epileptic zone~\cite{PRE:Andrzejak}. The wide dispersion in H for the EEG recorded from the epileptic zone in seizure free intervals ($\star$)  indicates that the epileptiform discharges are intermittent probably due to the chronic presence of abnormal epileptogenic tissues~\cite{Epilepsia:Schwartzkroin,Bromfield}. EEG recording during seizure may not be economical and hence it may be better to locate epileptic zone by recording EEG during seizure free intervals. These analyses show the possibility of detecting the onset of the seizure state from the time dependent Hurst exponent estimated during the transition from normal to seizure state~\cite{physicaA:Cajueiro}. Fig~\ref{fig:all_H} shows the Hurst exponents for all the epochs of five different sets that have been discussed above.

\begin{figure}
\centering
\includegraphics[height=6cm]{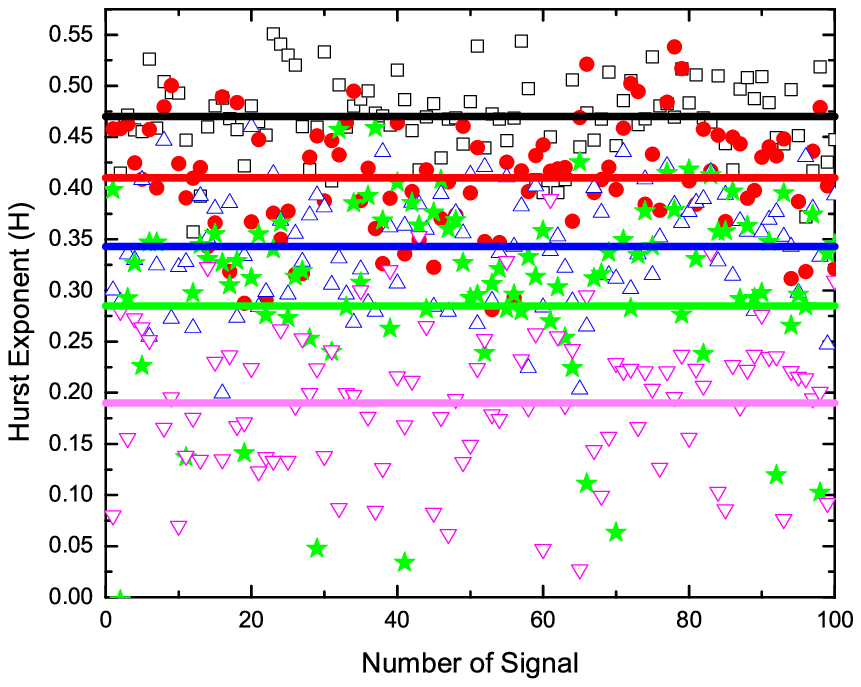}
\caption{Hurst exponent for hundred time series and average H are represented by:
$\Box$ and $-$  for set A; $\textcolor[rgb]{1.00,0.00,0.00}{\bullet}$and $\textcolor[rgb]{1.00,0.00,0.00}{-}$ for set B; $\textcolor[rgb]{0.00,0.00,1.00}{\triangle}$ and $\textcolor[rgb]{0.00,0.00,1.00}{-}$ for set C; $\textcolor[rgb]{0,1,0}{\star}$ and $\textcolor[rgb]{0,1,0}{-}$ for set D;$\textcolor[rgb]{1.00,0.50,0.75}{\nabla}$ and $\textcolor[rgb]{1.00,0.50,1.00}{-}$  for set E.}
\label{fig:all_H}       
\end{figure}
\section{Conclusion}
\label{sec:Conclusion}

In this paper we have reinvestigated the EEG data of normal and epileptic subjects to get an insight into the brain dynamics at different imposed and diseased conditions using RS surrogate analysis, PDF and H exponents. From these analysis we have found that RS and PDF may be useful to find a broad difference between normal and epileptic subjects but not helpful for constrained and seizure free intervals. Whereas, using H exponent, we have obtained differences in characteristics for normal subjects with eyes open and closed, and epileptic subjects during seizure and seizure free interval. The H shows that the brain activity of a normal man is uncorrelated in nature whereas, epileptic brains show long range anticorrelation.




{\ifthenelse{\boolean{publ}}{\footnotesize}{\small}
 \bibliographystyle{bmc_article}  
  \bibliography{bmc_bib} }     


\begin{thebibliography}{10}
\providecommand{\url}[1]{[#1]}
\providecommand{\urlprefix}{}

\bibitem{PRE:Andrzejak}
Andrzejak RG, Lehnertz K, Mormann F, Rieke C, David P, Christian EE:
  \textbf{Indications of nonlinear deterministic and finite dimensional
  structures in time series of brain electrical activity: dependence on
  recording region and brain state}. \emph{Phys Rev E} 2001,
  \textbf{64}:061907.

\bibitem{PRE:Diambra}
Diambra L, Malta CP: \textbf{BRCA1 protein products: functional motifs}.
  \emph{Phys.Rev.E} 1999, \textbf{59}:929.

\bibitem{PLA:Rombouts}
Rombouts SARB, Keunen RWM, Stam CJ: \textbf{Investigation of nonlinear
  structure in multichannel EEG}. \emph{Phys. Letts. A} 1995, \textbf{202}:352.

\bibitem{chaos:Mathew}
Dafilis MP, David TJL, Cadusch PJ: \textbf{Robust chaos in a model of the
  electroencephalogram: Implications for brain dynamics}. \emph{Chaos} 2001,
  \textbf{11}:474.

\bibitem{PRE:Robinson}
Robinson PA, Rennie CJ, Rowe DL: \textbf{Dynamics of large-scale brain activity
  in normal arousal states and epileptic seizures}. \emph{Phys. Rev. E} 2002,
  \textbf{65}:041924.

\bibitem{ClinicalNeuro:stam}
Stam CJ: \textbf{Nonlinear dynamical analysis of EEG and MEG: Review of an
  emerging field}. \emph{Clinical Neurophysiology} 2005,
  \textbf{116}:2266--2301.

\bibitem{PRE:pradhan}
Pradhan N, Sadasivan PK: \textbf{Relevence of surrogate-data testing in
  electroencephalograms}. \emph{Phys Rev E} 1996, \textbf{53}:2684.

\bibitem{Neurocomputing:Anna}
Krakovsk\'{a} A, $\check{S}$tolc SJ: \textbf{Spectral decay vs. correlation
  dimension of EEG}. \emph{Neurocomputing} 2008, \textbf{71}:2978.

\bibitem{NeurosciLett:Pereda}
Pereda E, Gamundi A, Rial R, Gonzalez J: \textbf{Non-linear behaviour of human
  EEG: fractal exponent versus correlation dimension in awake and sleep
  stages}. \emph{Neurosci Lett} 1998, \textbf{250}:91.

\bibitem{NeurosciLett:Pereda1}
Pereda E, Gamundi A, Nicolau MC, Rial R, Gonzalez J: \textbf{Interhemispheric
  differences in awake and sleep human EEG: a comparison between nonlinear and
  spectral measures}. \emph{Neurosci Lett} 1999, \textbf{263}:37.

\bibitem{EpilepsyBehavior:Hughes}
Hughes JR: \textbf{Progress in predicting seizure episodes with nonlinear
  methods}. \emph{Epilepsy \& Behavior} 2008, \textbf{12}:128.

\bibitem{ClinNeurophysiol:Mormann}
Mormann F, Kreuz T, Rieke C, Andrzejak RG, Kraskov A, David P, Christian~Elger
  E, Lehnertza K: \textbf{On the predictability of epileptic seizures}.
  \emph{Clin Neurophysiol} 2005, \textbf{116}:569.

\bibitem{JClinNeurophysiol:Jerger}
Jerger KK, Theoden NI, Joseph TF, Timothy S, Pecora L, Steven WL, Steven SJ:
  \textbf{Early seizure detection}. \emph{J Clin Neurophysiol} 2001,
  \textbf{18}:259.

\bibitem{IEEETransBiomedEng:Lopes}
Lopes~da Silva FH, Blanes W, Kalitzin SN, Parra J, Suffczynski P, Velis DN:
  \textbf{Dynamical diseases of brain systems: different routes to epileptic
  seizures}. \emph{IEEE Trans Biomed Eng} 2003, \textbf{50}:540.

\bibitem{AnnBiomedEng:Slutzky}
Slutzky MW, Cvitanovic P, Mogul DJ: \textbf{Deterministic chaos and noise in
  three in vitro hippocampal models of epilepsy}. \emph{Ann Biomed Eng} 2001,
  \textbf{29}:607.

\bibitem{JClinNeurophysiol:Savit}
Savit R, Li D, Zhou W, I D: \textbf{Understanding dynamic state changes in
  temporal lobe epilepsy.} \emph{J Clin Neurophysiol} 2001, \textbf{18}:246.

\bibitem{pre:Temujin}
Gautama T, Mandic DP, Hulle MMV: \textbf{Indications of nonlinear structures in
  brain electrical activity}. \emph{Phys. Rev. E} 2003, \textbf{67}:046204.

\bibitem{physicaD:Harikrishnan}
Harikrishnan K, Misra R, Ambika G, Kembhavi A: \textbf{A non-subjective
  approach to the GP algorithm for analysing noisy time series}. \emph{Physica
  D} 2006, \textbf{215}:137.

\bibitem{web:eeg}
\textbf{EEG time series download page}
  \urlprefix\url{[http://www.meb.uni-bonn.de/epileptologie/cms/front_content.p%
hp?idcat=193]}.

\bibitem{BrainTopogr:Pijn}
Pijn JP, Velis DN, van~der Heyden MJ, DeGoede J, van Veelen CW, Lopes~da Silva
  FH: \textbf{Nonlinear dynamics of epileptic seizures on basis of intracranial
  EEG recordings}. \emph{Brain Topogr} 1997, \textbf{9}:249.

\bibitem{ClinicalNeurophysiology:shen}
Shen Y, Olbricha E, Achermann P, Meier P: \textbf{Dimensional complexity and
  spectral properties of the human sleep EEG}. \emph{Clinical Neurophysiology}
  2003, \textbf{114}:199.

\bibitem{ClinicalNeurophysiology:Lee}
Lee JS, Yang BH, Lee JH, Choi JH, Choi IG, Kim SB: \textbf{Detrended
  fluctuation analysis of resting EEG in depressed outpatients and healthy
  controls}. \emph{Clinical Neurophysiology} 2007, \textbf{118}:2489.

\bibitem{ComputerMethodsandProgramsinBiomedicine:kannathal}
Kannathal N, Acharya UR, Lim C, Sadasivan PK: \textbf{Characterization of
  EEG$-$A comparative study}. \emph{Computer Methods and Programs in
  Biomedicine} 2005, \textbf{80}:17.

\bibitem{ClinicalNeurophysiology:Freeman}
Freeman WJ, Holmes MD, Burke BC, Vanhatalo S: \textbf{Spatial spectra of scalp
  EEG and EMG from awake humans}. \emph{Clinical Neurophysiology} 2003,
  \textbf{114}:1053.

\bibitem{pre:robinson1}
Robinson PA, Rennie CJ, Wright JJ, Bahramali H, Gordon E, Rowe DL:
  \textbf{Prediction of electroencephalographic spectra from neurophysiology}.
  \emph{Phys. Rev.E} 2001, \textbf{63}:021903.

\bibitem{JournalNeuroscienceMethods:nikolic}
Nikolic D, Moca VV, Singera W, Muresan RC: \textbf{Properties of multivariate
  data investigated by fractal dimensionality}. \emph{Journal of Neuroscience
  Methods} 2008, \textbf{172}:27.

\bibitem{BioMedicalEngineeringOnLine:Kannathal1}
Natarajan K, Acharya R, Alias F, Tiboleng T, Puthusserypady SK:
  \textbf{Nonlinear analysis of EEG signals at different mental states}.
  \emph{BioMedical Engineering OnLine} 2004, \textbf{3}:7.

\bibitem{physicaD:theiler}
Theiler J, Eubank S, Longtin A, Galdrikian B, Farmer D: \textbf{Testing for
  nonlinearity in time series: the method of surrogate data}. \emph{Physica D}
  1992, \textbf{58}:77.

\bibitem{Bifur:Nakamura}
Nakamura T, Small M: \textbf{Applying the method of Small-Shuffle surrogate
  data: Testing for dynamics in fluctuating data with trends}.
  \emph{International Journal of Bifurcations and Chaos} 2006,
  \textbf{16}:3581.

\bibitem{ieee:small}
Small M, Tse CK: \textbf{Detecting determinism in time series: The method of
  surrogate data}. \emph{IEEE Transactions on circuits and systems-I} 2003,
  \textbf{50}:663.

\bibitem{POP:carreras1}
van Milligen BP, Pedrosa MA, Balb\'{i}n R, Hidalgo C, Newman DE, S\'{a}nchez E,
  Frances M, Garc\'{i}a-Cort\'{e}s I, Bleuel J, Endler M, Riccardi C, Davies S,
  Matthews GF, Latten A, Klinger T: \textbf{Self-similarity of the plasma edge
  fluctuations}. \emph{Phys. Plasmas} 1998, \textbf{5}:3632.

\bibitem{chaos:dori}
Dori G, Fishman S, Ben-Haim SA: \textbf{The correlation dimension of rat hearts
  in an experimentally controlled environment}. \emph{Chaos} 2000,
  \textbf{10}:257.

\bibitem{chaos:Harrison}
Harrison MAF, Osorio I, Frei MG, Asuri S, Lai YC: \textbf{Correlation dimension
  and integral do not predict epileptic seizures}. \emph{Chaos} 2005,
  \textbf{15}:033106.

\bibitem{PhyslettA:jaman}
Nurujjaman M, Iyengar ANS: \textbf{Realization of SOC behavior in a dc glow
  discharge plasma}. \emph{Phys. Letts. A} 2007, \textbf{360}:717.

\bibitem{physicaA:Cajueiro}
Cajueiro DO, Tabak BM: \textbf{The Hurst exponent over time: testing the
  assertion that emerging markets are becoming more efficient}. \emph{Physica
  A} 2004, \textbf{336}:521.

\bibitem{Epilepsia:Schwartzkroin}
Schwartzkroin PA: \textbf{Progress in Epilepsy Research: Origins of the
  Epileptic State 38 (8): 853-858 (1997)}. \emph{Epilepsia} 1997,
  \textbf{38}:853.

\bibitem{Bromfield}
Bromfield EB: \textbf{Epileptiform Discharges}
  \urlprefix\url{[http://emedicine.medscape.com/article/1138880-overview]}.

\end{thebibliography}

\newcommand{\BMCxmlcomment}[1]{}

\BMCxmlcomment{

<refgrp>

<bibl id="B1">
  <title><p>Indications of nonlinear deterministic and finite dimensional
  structures in time series of brain electrical activity: dependence on
  recording region and brain state</p></title>
  <aug>
    <au><snm>Andrzejak</snm><fnm>RG</fnm></au>
    <au><snm>Lehnertz</snm><fnm>K</fnm></au>
    <au><snm>Mormann</snm><fnm>F</fnm></au>
    <au><snm>Rieke</snm><fnm>C</fnm></au>
    <au><snm>David</snm><fnm>P</fnm></au>
    <au><snm>Christian</snm><fnm>EE</fnm></au>
  </aug>
  <source>Phys Rev E</source>
  <pubdate>2001</pubdate>
  <volume>64</volume>
  <fpage>061907</fpage>
</bibl>

<bibl id="B2">
  <title><p>BRCA1 protein products: functional motifs</p></title>
  <aug>
    <au><snm>Diambra</snm><fnm>L</fnm></au>
    <au><snm>Malta</snm><fnm>C P</fnm></au>
  </aug>
  <source>Phys.Rev.E</source>
  <pubdate>1999</pubdate>
  <volume>59</volume>
  <fpage>929</fpage>
</bibl>

<bibl id="B3">
  <title><p>Investigation of nonlinear structure in multichannel
  EEG</p></title>
  <aug>
    <au><snm>Rombouts</snm><fnm>S A R B</fnm></au>
    <au><snm>Keunen</snm><fnm>R W M</fnm></au>
    <au><snm>Stam</snm><fnm>C J</fnm></au>
  </aug>
  <source>Phys. Letts. A</source>
  <pubdate>1995</pubdate>
  <volume>202</volume>
  <fpage>352</fpage>
</bibl>

<bibl id="B4">
  <title><p>Robust chaos in a model of the electroencephalogram: Implications
  for brain dynamics</p></title>
  <aug>
    <au><snm>Dafilis</snm><fnm>MP</fnm></au>
    <au><snm>David</snm><fnm>TJL</fnm></au>
    <au><snm>Cadusch</snm><fnm>PJ</fnm></au>
  </aug>
  <source>Chaos</source>
  <pubdate>2001</pubdate>
  <volume>11</volume>
  <fpage>474</fpage>
</bibl>

<bibl id="B5">
  <title><p>Dynamics of large-scale brain activity in normal arousal states and
  epileptic seizures</p></title>
  <aug>
    <au><snm>Robinson</snm><fnm>P. A.</fnm></au>
    <au><snm>Rennie</snm><fnm>C. J.</fnm></au>
    <au><snm>Rowe</snm><fnm>D. L.</fnm></au>
  </aug>
  <source>Phys. Rev. E</source>
  <pubdate>2002</pubdate>
  <volume>65</volume>
  <fpage>041924</fpage>
</bibl>

<bibl id="B6">
  <title><p>Nonlinear dynamical analysis of EEG and MEG: Review of an emerging
  field</p></title>
  <aug>
    <au><snm>Stam</snm><fnm>C J</fnm></au>
  </aug>
  <source>Clinical Neurophysiology</source>
  <pubdate>2005</pubdate>
  <volume>116</volume>
  <fpage>2266</fpage>
  <lpage>2301</lpage>
</bibl>

<bibl id="B7">
  <title><p>Relevence of surrogate-data testing in
  electroencephalograms</p></title>
  <aug>
    <au><snm>Pradhan</snm><fnm>N</fnm></au>
    <au><snm>Sadasivan</snm><fnm>P. K.</fnm></au>
  </aug>
  <source>Phys Rev E</source>
  <pubdate>1996</pubdate>
  <volume>53</volume>
  <fpage>2684</fpage>
</bibl>

<bibl id="B8">
  <title><p>Spectral decay vs. correlation dimension of EEG</p></title>
  <aug>
    <au><snm>Krakovsk\'{a}</snm><fnm>A</fnm></au>
    <au><snm>$\check{S}$tolc</snm><fnm>SJ</fnm></au>
  </aug>
  <source>Neurocomputing</source>
  <pubdate>2008</pubdate>
  <volume>71</volume>
  <fpage>2978</fpage>
</bibl>

<bibl id="B9">
  <title><p>Non-linear behaviour of human EEG: fractal exponent versus
  correlation dimension in awake and sleep stages</p></title>
  <aug>
    <au><snm>Pereda</snm><fnm>E</fnm></au>
    <au><snm>Gamundi</snm><fnm>A</fnm></au>
    <au><snm>Rial</snm><fnm>R</fnm></au>
    <au><snm>Gonzalez</snm><fnm>J.</fnm></au>
  </aug>
  <source>Neurosci Lett</source>
  <pubdate>1998</pubdate>
  <volume>250</volume>
  <fpage>91</fpage>
</bibl>

<bibl id="B10">
  <title><p>Interhemispheric differences in awake and sleep human EEG: a
  comparison between nonlinear and spectral measures</p></title>
  <aug>
    <au><snm>Pereda</snm><fnm>E</fnm></au>
    <au><snm>Gamundi</snm><fnm>A</fnm></au>
    <au><snm>Nicolau</snm><fnm>M C</fnm></au>
    <au><snm>Rial</snm><fnm>R</fnm></au>
    <au><snm>Gonzalez</snm><fnm>J.</fnm></au>
  </aug>
  <source>Neurosci Lett</source>
  <pubdate>1999</pubdate>
  <volume>263</volume>
  <fpage>37</fpage>
</bibl>

<bibl id="B11">
  <title><p>Progress in predicting seizure episodes with nonlinear
  methods</p></title>
  <aug>
    <au><snm>Hughes</snm><fnm>JR</fnm></au>
  </aug>
  <source>Epilepsy \& Behavior</source>
  <pubdate>2008</pubdate>
  <volume>12</volume>
  <fpage>128</fpage>
</bibl>

<bibl id="B12">
  <title><p>On the predictability of epileptic seizures</p></title>
  <aug>
    <au><snm>Mormann</snm><fnm>F</fnm></au>
    <au><snm>Kreuz</snm><fnm>T</fnm></au>
    <au><snm>Rieke</snm><fnm>C</fnm></au>
    <au><snm>Andrzejak</snm><fnm>RG</fnm></au>
    <au><snm>Kraskov</snm><fnm>A</fnm></au>
    <au><snm>David</snm><fnm>P</fnm></au>
    <au><snm>Christian Elger</snm><fnm>E.</fnm></au>
    <au><snm>Lehnertza</snm><fnm>K</fnm></au>
  </aug>
  <source>Clin Neurophysiol</source>
  <pubdate>2005</pubdate>
  <volume>116</volume>
  <fpage>569</fpage>
</bibl>

<bibl id="B13">
  <title><p>Early seizure detection</p></title>
  <aug>
    <au><snm>Jerger</snm><fnm>KK</fnm></au>
    <au><snm>Theoden</snm><fnm>NI</fnm></au>
    <au><snm>Joseph</snm><fnm>TF</fnm></au>
    <au><snm>Timothy</snm><fnm>S</fnm></au>
    <au><snm>Pecora</snm><fnm>L</fnm></au>
    <au><snm>Steven</snm><fnm>WL</fnm></au>
    <au><snm>Steven</snm><fnm>SJ</fnm></au>
  </aug>
  <source>J Clin Neurophysiol</source>
  <pubdate>2001</pubdate>
  <volume>18</volume>
  <fpage>259</fpage>
</bibl>

<bibl id="B14">
  <title><p>Dynamical diseases of brain systems: different routes to epileptic
  seizures</p></title>
  <aug>
    <au><snm>Silva</snm><fnm>F H</fnm></au>
    <au><snm>Blanes</snm><fnm>W</fnm></au>
    <au><snm>Kalitzin</snm><fnm>S N</fnm></au>
    <au><snm>Parra</snm><fnm>J</fnm></au>
    <au><snm>Suffczynski</snm><fnm>P</fnm></au>
    <au><snm>Velis</snm><fnm>D N</fnm></au>
  </aug>
  <source>IEEE Trans Biomed Eng</source>
  <pubdate>2003</pubdate>
  <volume>50</volume>
  <fpage>540</fpage>
</bibl>

<bibl id="B15">
  <title><p>Deterministic chaos and noise in three in vitro hippocampal models
  of epilepsy</p></title>
  <aug>
    <au><snm>Slutzky</snm><fnm>M W</fnm></au>
    <au><snm>Cvitanovic</snm><fnm>P</fnm></au>
    <au><snm>Mogul</snm><fnm>D J</fnm></au>
  </aug>
  <source>Ann Biomed Eng</source>
  <pubdate>2001</pubdate>
  <volume>29</volume>
  <fpage>607</fpage>
</bibl>

<bibl id="B16">
  <title><p>Understanding dynamic state changes in temporal lobe
  epilepsy.</p></title>
  <aug>
    <au><snm>Savit</snm><fnm>R</fnm></au>
    <au><snm>Li</snm><fnm>D</fnm></au>
    <au><snm>Zhou</snm><fnm>W</fnm></au>
    <au><snm>I.</snm><fnm>D</fnm></au>
  </aug>
  <source>J Clin Neurophysiol</source>
  <pubdate>2001</pubdate>
  <volume>18</volume>
  <fpage>246</fpage>
</bibl>

<bibl id="B17">
  <title><p>Indications of nonlinear structures in brain electrical
  activity</p></title>
  <aug>
    <au><snm>Gautama</snm><fnm>T</fnm></au>
    <au><snm>Mandic</snm><fnm>DP</fnm></au>
    <au><snm>Hulle</snm><fnm>MMV</fnm></au>
  </aug>
  <source>Phys. Rev. E</source>
  <pubdate>2003</pubdate>
  <volume>67</volume>
  <fpage>046204</fpage>
</bibl>

<bibl id="B18">
  <title><p>A non-subjective approach to the GP algorithm for analysing noisy
  time series</p></title>
  <aug>
    <au><snm>Harikrishnan</snm><fnm>KP</fnm></au>
    <au><snm>Misra</snm><fnm>R</fnm></au>
    <au><snm>Ambika</snm><fnm>G</fnm></au>
    <au><snm>Kembhavi</snm><fnm>AK</fnm></au>
  </aug>
  <source>Physica D</source>
  <pubdate>2006</pubdate>
  <volume>215</volume>
  <fpage>137</fpage>
</bibl>

<bibl id="B19">
  <title><p>EEG time series download page</p></title>
  <url>http://www.meb.uni-bonn.de/epileptologie/cms/front_content.php?idcat=19%
3</url>
</bibl>

<bibl id="B20">
  <title><p>Nonlinear dynamics of epileptic seizures on basis of intracranial
  EEG recordings</p></title>
  <aug>
    <au><snm>Pijn</snm><fnm>J P</fnm></au>
    <au><snm>Velis</snm><fnm>D N</fnm></au>
    <au><snm>Heyden</snm><fnm>M J</fnm></au>
    <au><snm>DeGoede</snm><fnm>J</fnm></au>
    <au><snm>Veelen</snm><fnm>C W</fnm></au>
    <au><snm>Silva</snm><fnm>F H</fnm></au>
  </aug>
  <source>Brain Topogr</source>
  <pubdate>1997</pubdate>
  <volume>9</volume>
  <fpage>249</fpage>
</bibl>

<bibl id="B21">
  <title><p>Dimensional complexity and spectral properties of the human sleep
  EEG</p></title>
  <aug>
    <au><snm>Shen</snm><fnm>Y.</fnm></au>
    <au><snm>Olbricha</snm><fnm>E.</fnm></au>
    <au><snm>Achermann</snm><fnm>P.</fnm></au>
    <au><snm>Meier</snm><fnm>P.F.</fnm></au>
  </aug>
  <source>Clinical Neurophysiology</source>
  <pubdate>2003</pubdate>
  <volume>114</volume>
  <fpage>199</fpage>
</bibl>

<bibl id="B22">
  <title><p>Detrended fluctuation analysis of resting EEG in depressed
  outpatients and healthy controls</p></title>
  <aug>
    <au><snm>Lee</snm><fnm>JS</fnm></au>
    <au><snm>Yang</snm><fnm>BH</fnm></au>
    <au><snm>Lee</snm><fnm>JH</fnm></au>
    <au><snm>Choi</snm><fnm>JH</fnm></au>
    <au><snm>Choi</snm><fnm>IG</fnm></au>
    <au><snm>Kim</snm><fnm>SB</fnm></au>
  </aug>
  <source>Clinical Neurophysiology</source>
  <pubdate>2007</pubdate>
  <volume>118</volume>
  <fpage>2489</fpage>
</bibl>

<bibl id="B23">
  <title><p>Characterization of EEG$-$A comparative study</p></title>
  <aug>
    <au><snm>Kannathal</snm><fnm>N.</fnm></au>
    <au><snm>Acharya</snm><fnm>UR</fnm></au>
    <au><snm>Lim</snm><fnm>C.M.</fnm></au>
    <au><snm>Sadasivan</snm><fnm>P .K.</fnm></au>
  </aug>
  <source>Computer Methods and Programs in Biomedicine</source>
  <pubdate>2005</pubdate>
  <volume>80</volume>
  <fpage>17</fpage>
</bibl>

<bibl id="B24">
  <title><p>Spatial spectra of scalp EEG and EMG from awake humans</p></title>
  <aug>
    <au><snm>Freeman</snm><fnm>WJ</fnm></au>
    <au><snm>Holmes</snm><fnm>MD</fnm></au>
    <au><snm>Burke</snm><fnm>BC</fnm></au>
    <au><snm>Vanhatalo</snm><fnm>S</fnm></au>
  </aug>
  <source>Clinical Neurophysiology</source>
  <pubdate>2003</pubdate>
  <volume>114</volume>
  <fpage>1053</fpage>
</bibl>

<bibl id="B25">
  <title><p>Prediction of electroencephalographic spectra from
  neurophysiology</p></title>
  <aug>
    <au><snm>Robinson</snm><fnm>P. A.</fnm></au>
    <au><snm>Rennie</snm><fnm>C. J.</fnm></au>
    <au><snm>Wright</snm><fnm>J. J.</fnm></au>
    <au><snm>Bahramali</snm><fnm>H.</fnm></au>
    <au><snm>Gordon</snm><fnm>E.</fnm></au>
    <au><snm>Rowe</snm><fnm>D. L.</fnm></au>
  </aug>
  <source>Phys. Rev.E</source>
  <pubdate>2001</pubdate>
  <volume>63</volume>
  <fpage>021903</fpage>
</bibl>

<bibl id="B26">
  <title><p>Properties of multivariate data investigated by fractal
  dimensionality</p></title>
  <aug>
    <au><snm>Nikolic</snm><fnm>D</fnm></au>
    <au><snm>Moca</snm><fnm>VV</fnm></au>
    <au><snm>Singera</snm><fnm>W</fnm></au>
    <au><snm>Muresan</snm><fnm>RC</fnm></au>
  </aug>
  <source>Journal of Neuroscience Methods</source>
  <pubdate>2008</pubdate>
  <volume>172</volume>
  <fpage>27</fpage>
</bibl>

<bibl id="B27">
  <title><p>Nonlinear analysis of EEG signals at different mental
  states</p></title>
  <aug>
    <au><snm>Natarajan</snm><fnm>K</fnm></au>
    <au><snm>Acharya</snm><fnm>R</fnm></au>
    <au><snm>Alias</snm><fnm>F</fnm></au>
    <au><snm>Tiboleng</snm><fnm>T</fnm></au>
    <au><snm>Puthusserypady</snm><fnm>SK</fnm></au>
  </aug>
  <source>BioMedical Engineering OnLine</source>
  <pubdate>2004</pubdate>
  <volume>3</volume>
  <fpage>7</fpage>
</bibl>

<bibl id="B28">
  <title><p>Testing for nonlinearity in time series: the method of surrogate
  data</p></title>
  <aug>
    <au><snm>Theiler</snm><fnm>J</fnm></au>
    <au><snm>Eubank</snm><fnm>S</fnm></au>
    <au><snm>Longtin</snm><fnm>A</fnm></au>
    <au><snm>Galdrikian</snm><fnm>B</fnm></au>
    <au><snm>Farmer</snm><fnm>D</fnm></au>
  </aug>
  <source>Physica D</source>
  <pubdate>1992</pubdate>
  <volume>58</volume>
  <fpage>77</fpage>
</bibl>

<bibl id="B29">
  <title><p>Applying the method of Small-Shuffle surrogate data: Testing for
  dynamics in fluctuating data with trends</p></title>
  <aug>
    <au><snm>Nakamura</snm><fnm>T.</fnm></au>
    <au><snm>Small</snm><fnm>M.</fnm></au>
  </aug>
  <source>International Journal of Bifurcations and Chaos</source>
  <pubdate>2006</pubdate>
  <volume>16</volume>
  <fpage>3581</fpage>
</bibl>

<bibl id="B30">
  <title><p>Detecting determinism in time series: The method of surrogate
  data</p></title>
  <aug>
    <au><snm>Small</snm><fnm>M.</fnm></au>
    <au><snm>Tse</snm><fnm>CK</fnm></au>
  </aug>
  <source>IEEE Transactions on circuits and systems-I</source>
  <pubdate>2003</pubdate>
  <volume>50</volume>
  <fpage>663</fpage>
</bibl>

<bibl id="B31">
  <title><p>Self-similarity of the plasma edge fluctuations</p></title>
  <aug>
    <au><snm>Milligen</snm><fnm>BP</fnm></au>
    <au><snm>Pedrosa</snm><fnm>M. A.</fnm></au>
    <au><snm>Balb\'{i}n</snm><fnm>R.</fnm></au>
    <au><snm>Hidalgo</snm><fnm>C.</fnm></au>
    <au><snm>Newman</snm><fnm>D. E.</fnm></au>
    <au><snm>S\'{a}nchez</snm><fnm>E.</fnm></au>
    <au><snm>Frances</snm><fnm>M.</fnm></au>
    <au><snm>Garc\'{i}a Cort\'{e}s</snm><fnm>I.</fnm></au>
    <au><snm>Bleuel</snm><fnm>J.</fnm></au>
    <au><snm>Endler</snm><fnm>M.</fnm></au>
    <au><snm>Riccardi</snm><fnm>C.</fnm></au>
    <au><snm>Davies</snm><fnm>S.</fnm></au>
    <au><snm>Matthews</snm><fnm>G. F.</fnm></au>
    <au><snm>Latten</snm><fnm>A.</fnm></au>
    <au><snm>Klinger</snm><fnm>T.</fnm></au>
  </aug>
  <source>Phys. Plasmas</source>
  <pubdate>1998</pubdate>
  <volume>5</volume>
  <fpage>3632</fpage>
</bibl>

<bibl id="B32">
  <title><p>The correlation dimension of rat hearts in an experimentally
  controlled environment</p></title>
  <aug>
    <au><snm>Dori</snm><fnm>G</fnm></au>
    <au><snm>Fishman</snm><fnm>S</fnm></au>
    <au><snm>Ben Haim</snm><fnm>S. A.</fnm></au>
  </aug>
  <source>Chaos</source>
  <pubdate>2000</pubdate>
  <volume>10</volume>
  <fpage>257</fpage>
</bibl>

<bibl id="B33">
  <title><p>Correlation dimension and integral do not predict epileptic
  seizures</p></title>
  <aug>
    <au><snm>Harrison</snm><fnm>MAF</fnm></au>
    <au><snm>Osorio</snm><fnm>I</fnm></au>
    <au><snm>Frei</snm><fnm>MG</fnm></au>
    <au><snm>Asuri</snm><fnm>S</fnm></au>
    <au><snm>Lai</snm><fnm>YC</fnm></au>
  </aug>
  <source>Chaos</source>
  <pubdate>2005</pubdate>
  <volume>15</volume>
  <fpage>033106</fpage>
</bibl>

<bibl id="B34">
  <title><p>Realization of SOC behavior in a dc glow discharge
  plasma</p></title>
  <aug>
    <au><snm>Nurujjaman</snm><fnm>M</fnm></au>
    <au><snm>Iyengar</snm><fnm>ANS</fnm></au>
  </aug>
  <source>Phys. Letts. A</source>
  <pubdate>2007</pubdate>
  <volume>360</volume>
  <fpage>717</fpage>
</bibl>

<bibl id="B35">
  <title><p>The Hurst exponent over time: testing the assertion that emerging
  markets are becoming more efficient</p></title>
  <aug>
    <au><snm>Cajueiro</snm><fnm>DO</fnm></au>
    <au><snm>Tabak</snm><fnm>BM</fnm></au>
  </aug>
  <source>Physica A</source>
  <pubdate>2004</pubdate>
  <volume>336</volume>
  <fpage>521</fpage>
</bibl>

<bibl id="B36">
  <title><p>Progress in Epilepsy Research: Origins of the Epileptic State 38
  (8): 853-858 (1997)</p></title>
  <aug>
    <au><snm>Schwartzkroin</snm><fnm>PA</fnm></au>
  </aug>
  <source>Epilepsia</source>
  <pubdate>1997</pubdate>
  <volume>38</volume>
  <fpage>853</fpage>
</bibl>

<bibl id="B37">
  <title><p>Epileptiform Discharges</p></title>
  <aug>
    <au><snm>Bromfield</snm><fnm>EB</fnm></au>
  </aug>
  <url>http://emedicine.medscape.com/article/1138880-overview</url>
</bibl>

</refgrp>
} 

\ifthenelse{\boolean{publ}}{\end{multicols}}{}

\end{bmcformat}
\end{document}